\documentclass[%
 aip,
 amsmath,amssymb,
 reprint,%
]{revtex4-1}

\usepackage{graphicx}
\usepackage{dcolumn}
\usepackage{bm}

\usepackage[utf8]{inputenc}
\usepackage[T1]{fontenc}
\usepackage{mathptmx}
\usepackage{etoolbox}
\usepackage{xcolor}
\usepackage{CJK}

\newcommand{\vct}[1] {\boldsymbol{#1}}
\renewcommand{\d}{\mathrm{d}}

\newcommand{\di} {d_\text{i}}

\newcommand{\PiD} {\text{Pi-D}}

\makeatletter
\def\@email#1#2{%
 \endgroup
 \patchcmd{\titleblock@produce}
  {\frontmatter@RRAPformat}
  {\frontmatter@RRAPformat{\produce@RRAP{*#1\href{mailto:#2}{#2}}}\frontmatter@RRAPformat}
  {}{}
}%
\makeatother
\begin{document}
\begin{CJK*}{UTF8}{gbsn}

\title[Revisiting pressure-strain]{Revisiting compressible and incompressible pressure-strain interaction in kinetic plasma turbulence}

\author{Subash Adhikari}
\email{subash@udel.edu}
\affiliation{Department of Physics and Astronomy,
University of Delaware, Newark, DE 19716, USA}
\author{Yan Yang (杨艳)}
\affiliation{Department of Physics and Astronomy,
University of Delaware, Newark, DE 19716, USA}
\author{William H. Matthaeus}
\affiliation{Department of Physics and Astronomy,
University of Delaware, Newark, DE 19716, USA}

\begin{abstract}
In this study, we revisit the pressure-strain interaction in kinetic turbulence, and in particular 
we re-examine the 
decomposition of pressure-strain interaction 
into compressive and incompressive parts.
The pressure dilatation ingredient is 
clearly due to plasma compressions, but here using kinetic particle-in-cell (PIC) simulations of plasma turbulence, 
it is demonstrated that the 
the remaining anisotropic part, often called Pi-D,
also contains contributions due to compressive, non solenoidal velocities of the particle species.
The compressive Pi-D can play a significant role in systems with low plasma $\beta$ even if the system starts with small density variations. The compressive ingredient of Pi-D is found to be anticorrelated with both incompressive Pi-D and pressure dilatation. 
\end{abstract}
\maketitle

\section{Introduction}
\end{CJK*}

Investigation of dissipation mechanisms in collisionless plasma turbulence has been an active area of study for decades. However, in recent years it has become recognized that the conversion of macroscopic (fluid) motions into internal energy 
must pass through the channel denoted as {\it pressure-strain interaction}~\citep{yangEA17-PoP, YangEA-PRE-17}. 
Considerations of pressure-strain interaction have emphasized its distinct role, relative to, for example the electromagnetic work~\citep{Zenitani2011new, yang2024electron}, that can convert magnetic energy into species flow energy, or heat conduction, that transports internal energy across space. 
Instead it is a rigorous consequence of Vlasov Maxwell theory that 
pressure-strain interaction is the energy transfer channel that converts 
energy from flow to thermal degrees of freedom. 
In addition to the elementary decomposition that separates pressure dilatation and Pi-D (see below),
different types of decomposition have been 
proposed in~\citet{cassak2022pressure}.
Here 
we revisit the decomposition of the pressure-strain interaction in kinetic plasma, and explore how this changes our previous understanding of pressure-strain interaction, and in particular its elements that may be described as incompressive and compressive.

The pressure-strain interaction (P:S) for any plasma species $\alpha$ can be decomposed as follows:
\begin{equation}
\underbrace{-\left( \boldsymbol{P}_\alpha \cdot \nabla \right) \cdot \boldsymbol{u}_\alpha}_{\text{P:S}_{\alpha}} = \underbrace{-p_{\alpha} \nabla \cdot \boldsymbol{u}_{\alpha}}_{p\theta_{\alpha}} \underbrace{-\boldsymbol{\Pi}_{\alpha}:\boldsymbol{D}_{\alpha}}_{\text{Pi-D}_{\alpha}},
\label{eq:PS}
\end{equation}
where $p_{\alpha}=P_{\alpha,ii}/3$ is the scalar pressure, $\Pi_{\alpha,ij}=P_{\alpha,ij}-p_{\alpha}\delta_{ij}$ is the deviatoric part of the pressure tensor, $D_{ij,\alpha}=\left(\partial_i u_{j,\alpha} + \partial_j u_{i,\alpha}\right)/2-\left(\nabla \cdot \boldsymbol{u}_{\alpha}\right) \delta_{ij}/3$ is the traceless symmetric part of the velocity gradient with $\mathbf{u}_\alpha$ being the velocity of species $\alpha$. Note that $p\theta_{\alpha}$ and $\text{Pi-D}_{\alpha}$ denote the isotropic and anisotropic part of the pressure-strain interaction $\text{P:S}_\alpha$ respectively. This decomposition was previously referred to as the compressive $p\theta_\alpha$ and incompressive $\text{Pi-D}_\alpha$ ingredients of the pressure-strain interaction. 

It is clear that $p\theta_{\alpha}$ arises from compression. Following the standard nomenclature in neutral gas dynamics, we denote this term as the pressure-dilatation. 
Here we will identify and quantify other contributions to pressure-strain interaction due to 
plasma compression.  
{\it While it is 
not trivial to quantify the effects
of compression on the pressure tensor itself,} 
the path to separation of compressive effects due to the velocities is quite straightforward. 
One can perform a Helmholtz decomposition~\citep{arfken2011mathematical} and separate the velocity $\mathbf{u}_\alpha$ for each species $\alpha$ field into irrotational $\mathbf{u}^c$ (compressive) and solenoidal $\mathbf{u}^s$ components i.e. $\mathbf{u}=\mathbf{u}_\alpha^s+\mathbf{u}_\alpha^s$ such that $\nabla \cdot \mathbf{u}^c=0$ and $\nabla \times \mathbf{u}^c=0$. Now the traceless symmetric part of the velocity gradient $\boldsymbol{D}_{\alpha}$ can be
further decomposed into compressive and incompressive parts:
\begin{equation}
    \boldsymbol{D}_\alpha=\boldsymbol{D}_\alpha^c + \boldsymbol{D}_\alpha^s
\end{equation}
which immediately implies the 
decomposition
\begin{equation}
\text{Pi-D}_{\alpha} = \text{Pi-D}_{\alpha}^c+\text{Pi-D}_{\alpha}^s,
\label{eq:PiDdecompose}
\end{equation}
which makes it explicit that 
$\text{Pi-D}_{\alpha}$ consists of 
both an incompressive contribution $\text{Pi-D}_{\alpha}^s$
and a compressive contribution $\text{Pi-D}_{\alpha}^c$.

It is shown that the compression effect is present in not only the pressure-dilatation term but also the anisotropic part of the pressure-strain interaction $\text{Pi-D}_{\alpha}$, which merits more attention. 
We should note in passing that the same decomposition of flow velocities can be carried out in magnetohydrodynamics (MHD), resulting 
in an analogous contrast between dissipation due to compressive flows  and due to solenoidal flows 
\citep{sarkar1991analysis,DonzisJohn20,Li2023anomaly,john2021does}. Of course in the MHD 
gas dynamics case, the dissipation
functions are expressed in terms of
collisional closures, 
thus obscuring the underlying role of 
pressure strain. 
In the present paper, we 
proceed to 
examine, for the collisionless plasma case, 
the compressive and incompressive contributions of $\text{Pi-D}_{\alpha}$ for electrons ($\alpha=e$) and ions ($\alpha=i$) in more detail, 
with the aim to correct our previous understandings on $\text{Pi-D}_{\alpha}$.

\section{Simulations}\label{sec:data}
In this study we use
fully kinetic PIC simulations of turbulence employing the P3D code \citep{ZeilerJGR02}. The simulations are performed in a two dimensional X-Y plane with all three components of the field vectors but no variation along the out-of-plane Z direction.
In P3D, a standard set of normalization is used largely based on protons, with 
number density normalized to a reference value $n_r$,
mass normalized to proton mass $m_\text{i}$,
charge normalized to proton charge $e$, 
and
magnetic field normalized to a reference field $B_r$.
Length is normalized to the proton inertial length $\d_i$,
time normalized to the proton cyclotron time $\omega_{ci}^{-1}$,
and velocity is normalized to the 
  equivalent reference Alfv{\'e}n speed
$V_{Ar} = B_r/\left(\mu_0 m_\text{i} n_r\right)^{1/2}$.

The simulations are performed in a square periodic domain of size $L=150 \, \di$ with $4096^2$ spatial grid points.
For numerical expediency both simulations employ artificially low values of
the proton to electron mass ratio, $m_\text{i}/m_\text{e} = 25$,
and
the speed of light, $ c=15 \,V_{Ar}$. We use a total of 
$3200$ particles of each species
per cell ($\sim 10^{11}$ total particles).
The system is
a decaying initial value problem,
starting with uniform densities
and temperatures for both species.
A uniform magnetic field, $B_0 = 1.0$, is directed
normal to the plane, in the $Z$ direction.
The initial velocity $\vct{u}$ and magnetic $\vct{B}$ fluctuations are transverse to $B_0$
(corresponding in linear theory to a small amplitude ``Alfv\'en mode''), populating 
Fourier wavevectors $2\le |\vct{k}| \le 4$ with
random phases and prescribed amplitudes. The initial 
fluctuation energy, flow plus magnetic, is $E=0.1$, and the 
normalized cross helicity $\sigma_c$ is negligible. 
We vary the electron and proton temperatures to change the plasma beta, ratio of the thermal pressure to the magnetic pressure. Results from four simulations with ion and electron plasma beta $\beta_{i}=\beta_e=(0.03, 0.3, 0.6, 1.2)$ are presented in this study. For the $\beta=0.03$ case, the grid scale is reduced to half compared to the other cases to account for the reduced electron gyroscale. Also for convenience, we drop the subscript $i$ and $e$ with the understanding that $\beta$ represents plasma beta of individual species.

\section{Results}
\label{sc:results}
\begin{figure}
    \hspace{-0.75cm}
    \includegraphics[width=0.5\textwidth]{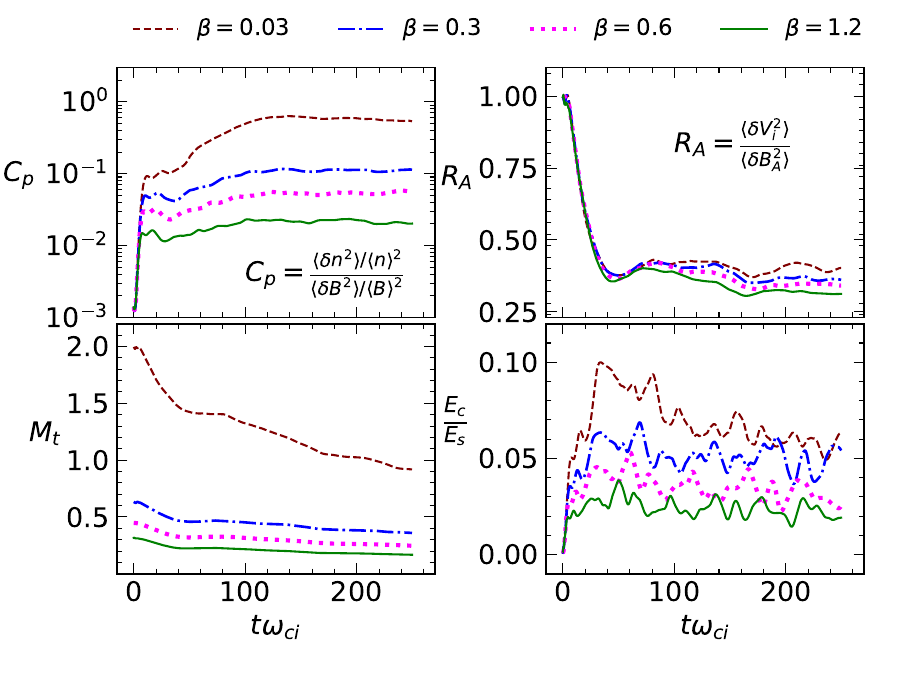}
    \caption{Time evolution of the plasma compressibility ratio $C_p$; Alf\'ven ratio $R_A$;  turbulent Mach number $M_t$;
    and the ratio of the total (ions plus electrons) compressible flow energy $E_c$ to solenoidal flow $E_s$ energy.}
    \label{fig:overview_plot}
\end{figure}

In Fig.~\ref{fig:overview_plot}a we show the time evolution of the plasma compressibility ratio 
$C_p=\big(\langle\delta n^2\rangle / \langle n\rangle^2 \big)/\big(\langle\delta B^2\rangle / \langle B\rangle^2 \big)$, where $\delta n$ is the root mean square (rms) density fluctuation and $\delta B$ is the rms fluctuation in the magnetic field. Note that all the simulations have been initialized with the same density fluctuation of $\delta n/\langle n \rangle=0.01178$.  For the simulation with 
the lowest $\beta$, the fluctuation $\delta n/\langle n \rangle$  
increases to $0.25$ as the system evolves.
In the highest $\beta$ case, $\delta n /\langle n \rangle$
increases to about $0.05$.
The measure of compressibility employed 
is through the compressibility ratio $C_P$ accounting for the possible anticorrelation between density and magnetic field~\citep{wang2016compressible}.
Initially all runs have $C_P=10^{-3}$, and this 
increases in time as turbulence develops. 
In particular, it is found that the 
compressibility ratio $C_P$ achieves larger 
values for simulations with smaller $\beta$ . 
For the $\beta=1.2$ case, $C_P$ saturates at about $2\times 10^{-2}$ while for $\beta=0.03$, $C_P$ increases to about $0.6$. This implies that, especially for low $\beta$ cases, even though the simulations start with an incompressible initial condition, compressibility effects can be significant as the system evolves. 

Fig.~\ref{fig:overview_plot}b shows the time evolution of the Alfv\'en ratio $R_A=\langle \delta V_i^2 \rangle /\langle \delta B_A^2\rangle $, where $\delta V_i$ is the ion-velocity fluctuation amplitude and $\delta B_A$ is the amplitude of the magnetic field ($\mathbf{B}$) fluctuation in Alfv\'en units i.e. $\mathbf{B}_A=\mathbf{B}/\sqrt{4\pi m_i n_i}$ with $m_i$ and $n_i$ being the mass and number density of ions respectively. Initially, the velocity and magnetic field fluctuation energies are equal. As the system evolves, 
the fluctuation in magnetic field dominates over the velocity field. While this ratio is almost independent of plasma $\beta$ until about $tw_{ci}=40$, at later times systems with larger $\beta$ values show smaller Alfv\'en ratios. 

Fig.~\ref{fig:overview_plot}c shows the time evolution of the turbulent Mach number $M_t =\delta u/c_s$, where $\delta u $ is the root mean square (r.m.s) ion velocity fluctuation and $c_s$ is the sound speed estimated using ion temperature. Among all the systems, only the lowest beta case $\beta=0.03$ has $M_t>1$ which falls off rapidly compared to the other cases and goes below unity at late times. Fig.~\ref{fig:overview_plot}d shows the evolution of the ratio of the total (ion plus electron) compressible to incompressible flow energy $E_c/E_s$ defined as $E_{c,\alpha}=\frac{1}{2}\langle \rho_\alpha \lvert \mathbf{u}_\alpha^{c}\rvert^2\rangle$ and $E_{s,\alpha}=\frac{1}{2}\langle \rho_\alpha \lvert \mathbf{u}_\alpha^{s}\rvert^2\rangle$ for each species $\alpha$ with mass density $\rho_\alpha$. This quantity is greatest throughout the run for the $\beta = 0.03$ case, and the ordering progresses to lower values for the respective higher 
$\beta$ cases, which is consistent with the trends in Fig.~\ref{fig:overview_plot}a and c.

\begin{figure}
    \hspace{-0.75cm}
    \includegraphics[width=0.5\textwidth]{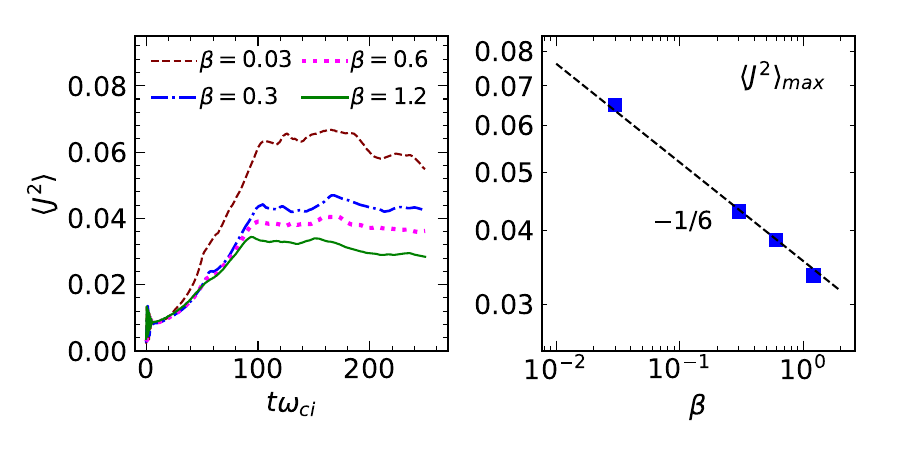}
    \caption{Left: Time evolution of the mean square current $\langle J^2 \rangle$ in all the runs. Right: Scaling the plateau of the maximum mean square current as a function of beta $\beta$. A line of slope $-1/6$ is drawn for reference.}
    \label{fig:jsq_plot}
\end{figure}

\begin{figure*}
    \centering
    \hspace{-1cm}
    \includegraphics[width=1\textwidth]{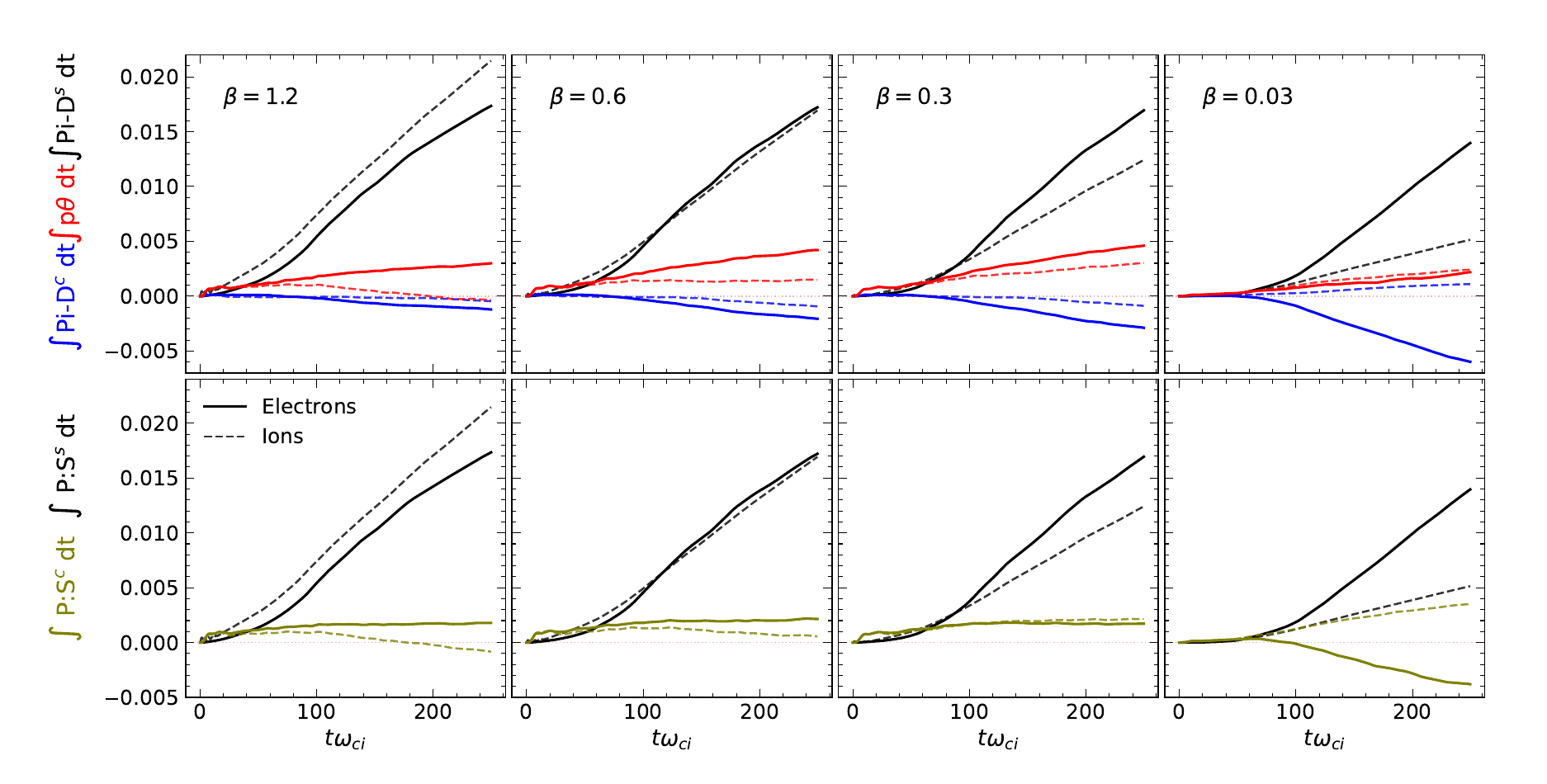}
    \caption{Top panel: Time evolution of the cumulative time integral of the compressible and incompressible contribution of $\text{Pi-D}$ along with the cumulative integral of $\text{p}\theta$ for both ions (dashed lines) and electrons (solid lines) for different plasma beta. Bottom Panel: Time evolution of the compressible and incompressible contributions to the pressure-strain interaction. Here, the sum of $\text{Pi-D}^c$ and $\text{p}\theta$ represents the compressible contribution to the pressure-strain interaction while $\text{Pi-D}^s$ represents the incompressible contribution.}
    \label{fig:pid_ptheta}
\end{figure*}

In Fig.~\ref{fig:jsq_plot}a we compare the time evolution of the mean square current $\langle J^2 \rangle$ in all four cases. Initially, the mean square current oscillates in all the simulations with the amplitude of the current decreasing with a higher value of $\beta$. 
This should be viewed as a startup transient~\citep{ParasharEA18}. 
Following 
$t\omega_{ci}=25$,
the mean square current for the $\beta=0.03$ case increases 
rapidly relative to other cases. 
The other simulations follow each other closely until $t\omega_{ci}=90$ before attaining slightly differing plateaus, once again ordered by their 
$\beta$ values.  The value of the mean square current around the plateau is shown in Fig.~\ref{fig:jsq_plot}b as a function of $\beta$. We find the empirical 
result that 
the maximum mean square current 
decreases with higher value of beta as
$\langle j^2 \rangle_{max} \sim \beta^{-1/6}$. 
We know of no obvious explanation for this 
empirical result.
\begin{figure}
    \includegraphics[width=0.45\textwidth]{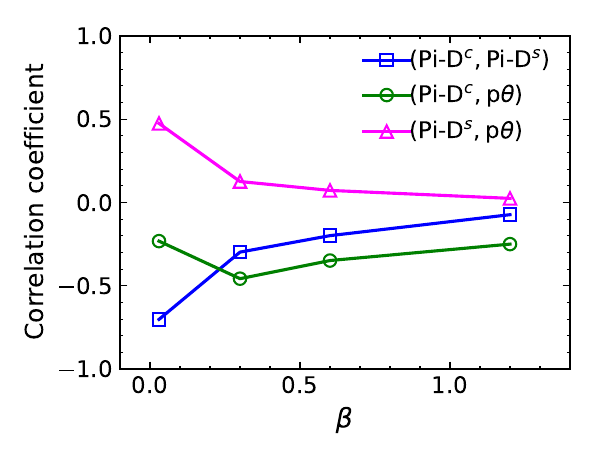}
    \caption{Correlation coefficient between(a) compressible and incompressible $\text{Pi-D}$ (blue squares), (b) compressible $\text{Pi-D}$ and $\text{p}\theta$(green circles), and (c) incompressible $\text{Pi-D}$ and $\text{p}\theta$ (magenta triangles) measured for electrons at $t\omega_{ci}=150.5$ as a function of $\beta$.}
    \label{fig:correlation_coeff}
\end{figure}

Next, in Fig.~\ref{fig:pid_ptheta} we examine the cumulative time integrals of pressure-strain interaction. This represents 
the net dissipation through 
the corresponding pathway to internal energy production. 
In the upper panel, we show the cumulative time integral of the solenoidal and compressible $\text{Pi-D}$ named as $\text{Pi-D}^s$ (black curve) and $\text{Pi-D}^c$ (blue curve), respectively.
The upper panels also show 
pressure dilatation, named as $\text{p}\theta$ (red curve) for both ions (dashed lines) and electrons (solid lines). 
For $\beta=1.2$ case, the time integral of $\text{Pi-D}^s_i>\text{Pi-D}^s_e$, and for $\beta=0.6$ case, the cumulative integrals $\text{Pi-D}^s_i\simeq \text{Pi-D}^s_e$.
For the lower $\beta$ cases, the time integral of $\text{Pi-D}^s_e$ dominates that of $\text{Pi-D}^s_i$. The integral of the compressible part of $\text{Pi-D}$ for ions is small and negative for all but $\beta=0.03$ case. However, the time integral of $\text{Pi-D}^c$ for electrons is always negative with the amplitude increasing with decreasing $\beta$. The amplitude of the time integral of $\text{Pi-D}^c$ for electrons is always larger than that of ions, in particular, the electron contribution dominates the ions for $\beta=0.03$. 

\begin{figure*}
    \centering
    \hspace{-1cm}
    \includegraphics[width=0.75\textwidth]{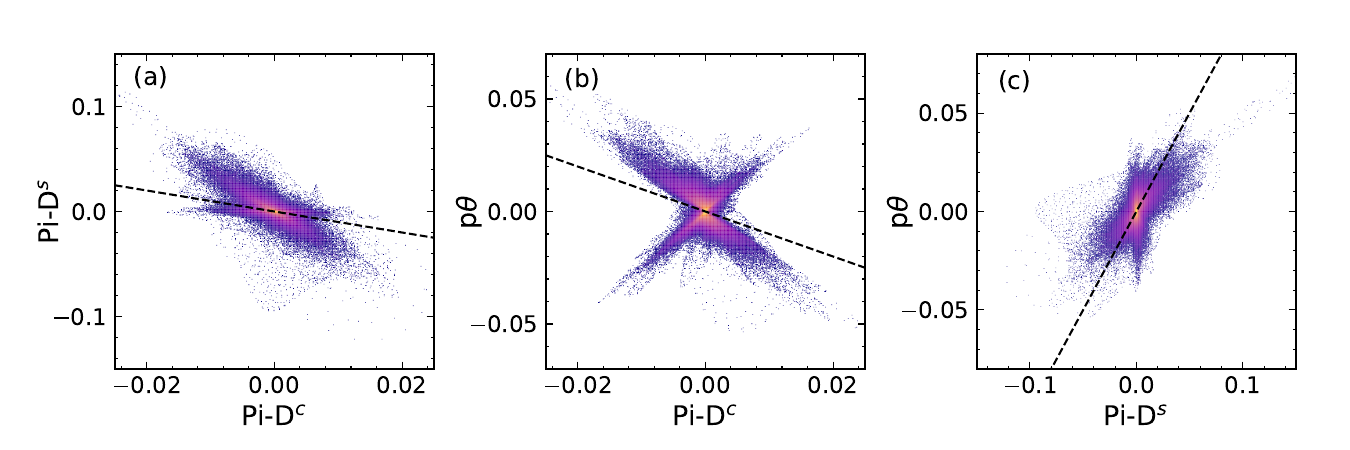}
    \caption{Joint pdf of (a) compressible and incompressible contributions to $\text{Pi-D}$ (left), (b) $\text{p}\theta$ and compressible $\text{Pi-D}$ (middle) and (c) $\text{p}\theta$ and incompressible $\text{Pi-D}$ for electrons in the system with $\beta=0.03$ at $t\omega_{ci}=150.5$. A line of slope $-1$ is drawn for reference in the left and middle panel while a line of slope $1$ is drawn on the rightmost plot.}
    \label{fig:joint_pdf}
\end{figure*}

\begin{figure*}
    \centering
    \hspace{-1cm}
    \includegraphics[width=0.75\textwidth]{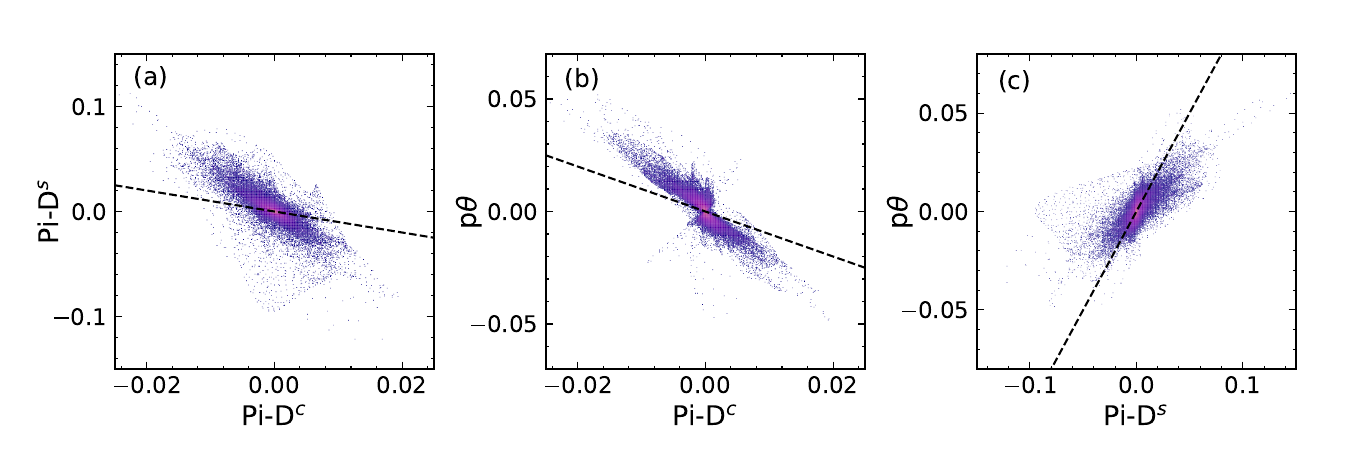}
    \caption{Same as Fig.~\ref{fig:joint_pdf} but conditioned over regions where the magnitude of the out-of-plane current density $j_z$ exceeds $0.75$.}
    \label{fig:joint_pdf_above0p75}
\end{figure*}

The well-known pressure dilatation also incorporates the compression effect. As shown in  Fig.~\ref{fig:pid_ptheta}, the cumulative integral of $\text{p}\theta$ for ions decreases as the system passes the maximum mean square current for $\beta=1.2$, stays almost flat for $\beta=0.6$, and increases for $\beta=0.3$. For the lowest value $\beta=0.03$, the integral of $p\theta_i$ 
continues to increase; 
however, its amplitude 
remains less than $p\theta_i$ for the $\beta=0.3$ case. 
Note that the time integral of $p\theta_e$ is larger than that of ions for $\beta=1.2,0.6,0.3$ with their differences decreasing with decreasing $\beta$. For $\beta=0.03$, the time integral of $p\theta_i$ exceeds that of electrons. 

Next, in the lower panel of Fig.~\ref{fig:pid_ptheta}, we combine the compressive contributions to the pressure-strain interaction i.e. the 
sum of 
$\text{Pi-D}^c + p\theta \equiv \text{P:S}^c$, and compare this with the time evolution of the incompressive contribution $\text{Pi-D}^s\equiv \text{P:S}^s$ for electrons and ions.
A subtle but potentially significant feature of the 
analysis in 
Fig.~\ref{fig:pid_ptheta}
is that the compressive contribution $\text{Pi-D}^c$ is 
always negative for electrons and almost always negative for protons, except at the highest $\beta$.
When this occurs, the compressive ``correction'' to $\text{Pi-D}$ is, in effect,
canceling part of the 
pressure dilatation. 

To further examine this 
phenomenon, in particular for electrons, 
in Fig.~\ref{fig:correlation_coeff}, we plot the correlation coefficient between different contributions to the electron pressure-strain interaction at $t\omega_{ci}=150.5$ as a function of $\beta$. Note that at this time all the systems are in the plateau of the mean square current (see Fig.~\ref{fig:jsq_plot}). As seen in Fig.~\ref{fig:correlation_coeff}, the compressible and incompressible parts of $\text{Pi-D}$ shown in blue squares
become more anticorrelated as $\beta$ decreases with $\beta=0.03$ case having a correlation coefficient of $r=-0.70$. While the correlation between $\text{Pi-D}^c$ and $\text{p}\theta$ (red circles) stays negative for all values of $\beta$, it decreases as $\beta$ decreases from $1.2$ to $0.3$, but for $\beta=0.03$ the correlation coefficient increases. The correlation between $\PiD^s$ and $\text{p}\theta$ (magenta triangles) is positive for all cases with the correlation increasing for decreasing $\beta$. Since the measure of the anticorrelation between $\text{Pi-D}^c$ and $\text{Pi-D}^s$, and the correlation between $\text{Pi-D}^s$ and $\text{p}\theta$ is maximum for the lowest $\beta$ case, we focus on this system and explore the associated probability distribution functions.

In Fig.~\ref{fig:joint_pdf}, we examine specifically  pairwise local correlations 
between several contributions to pressure strain for the $\beta=0.03$ case.
We plot joint probability distribution functions (pdf) 
for $\text{Pi-D}^s$ vs $\text{Pi-D}^c$, 
for $\text{p}\theta$ vs $\text{Pi-D}^c$, and for $\text{p}\theta$ vs $\text{Pi-D}^s$.
 For convenience, reference dashed lines of slope $-1$ and $+1$ are plotted. From Fig.~\ref{fig:joint_pdf}a, it is clear that the compressible and incompressible $\text{Pi-D}$ have a joint pdf favoring a negative correlation. For compressible $\text{Pi-D}$ and pressure dilatation, the joint pdf in Fig.~\ref{fig:joint_pdf}b shows mixed correlations.  Most of the points align to the line of anticorrelation, but there is a lesser population that exhibits positive correlations. 
 It is also clear from Fig.~\ref{fig:joint_pdf}c that the incompressible $\text{Pi-D}$ shows a very strong (positive) correlation with the pressure dilatation $\text{p}\theta$ even though there are some tails deviated from the perfect correlation line.

To understand these correlations better, we isolate the regions with strong out-of-plane current density $\lvert j_z \rvert/j_{rms} >0.75$ and Fig.~\ref{fig:joint_pdf_above0p75} shows the joint pdfs corresponding to these strong current regions. In this case, the arms-like extensions observed in Fig.~\ref{fig:joint_pdf} are largely absent. As a result, the correlation aligns better with the lines drawn for reference. For convenience, we compare the value of the correlation coefficients between these pressure-strain elements across regions 
with different intensities of the out-of-plane current ($j_z$) in Table~\ref{tab:tableI}. Therefore, in a system with low plasma $\beta$, within the localized current sheets, the compressible $\text{Pi-D}^c$ is anti-correlated to both the incompressible $\text{Pi-D}^s$ and pressure dilatation $\text{p}\theta$. Even though $\text{Pi-D}^c$ and $\text{p}\theta$, both are the compressive contributions to the pressure-strain interaction, the net effect of these quantities is 
found to be opposite to each other. 

\begin{table}
\fontsize{10}{11}\selectfont
\caption{\label{tab:tableI}Correlation coefficients between different contributions to the electron pressure-strain interaction conditioned on the amplitude of the normalized out-of-plane current $j_z$. }
\begin{ruledtabular}
\begin{tabular}{|c|c|c|c|}
 Regions & $\text{Pi-D}^c$, $\text{Pi-D}^s$ & $\text{Pi-D}^c$, $\text{p}\theta$ & $\text{Pi-D}^s$, $\text{p}\theta$ \\
 \hline
 $\lvert j_z \rvert >0.75$   & $-0.79$   & $-0.88$ &   $0.71$\\
 $\lvert j_z \rvert <0.75$&   $-0.69$  & $-0.16$   & $0.45$\\
 All & $-0.70$ & $-0.23$ &  $0.47$\\
\end{tabular}
\end{ruledtabular}
\end{table}

\section{Discussion and Conclusions}
\label{sec:disc}

In this brief report we
revisit the previously discussed decompositions 
of pressure-strain interaction into compressible and incompressible parts.
In earlier work, $\text{Pi-D}$, has been described, either explicitly or implicitly,
as the incompressible contribution to the pressure-strain interaction.
We report here the elementary, but possibly overlooked,
conclusion that when there is a nonuniform irrotational ingredient 
of the flow, then $\text{Pi-D}$ can include 
a compressible contribution hidden within it. 
The possibility for this to occur is seen directly from the 
Helmholtz decomposition of the velocities.

We examine this effect in plasma kinetic (PIC)
simulations by directly computing the 
compressible $\text{Pi-D}^c$ for electrons and ions. The diagnostics examined here have specifically emphasized the intriguing results for electrons. The finding is that 
the compressible $\text{Pi-D}^c$ can be significant in plasma simulations with low plasma $\beta$.
Further analysis
shows that 
compressible $\text{Pi-D}^c$ is anticorrelated to the incompressible $\text{Pi-D}^s$ and also 
is anticorrelated to pressure dilatation $\text{p}\theta$ when computed 
over the entire volume. 
But upon conditioning on current intensity, 
we find that this anticorrelation 
is most prominent in 
regions with strong currents.
Because of this anticorrelation, the net compressible contribution to the pressure-strain interaction may differ considerably from the estimate based on pressure dilatation alone. That is, especially for electrons, the 
newly characterized compressive $\text{Pi-D}^c$ opposes the effect of pressure 
dilatation. 
Recognition of this effect calls for 
revising estimates of the influence of pressure-strain on 
heating in compressive cases, relative to the standard 
incompressible heating estimates. Here we see that the globally averaged effect, at low plasma $\beta$ can be $50\%$ or more for electrons. The local effect, near stronger currents, may be much greater. 

As a final word, 
the potential implications of these results for interpretation of MMS observations are worth mentioning. 
For example, when one observes
$|p\theta|>|\text{Pi-D}|$
as is often the case in the magnetosheath~\citep{WangEA21-MMS,YangEA23}, 
it is not a necessary
conclusion that compressible effects greatly exceed incompressive dissipation effects. 
Rather, since we have observed that $\text{Pi-D}^c$ is often anticorrelated with $\text{Pi-D}^s$ and can be large, it may be masking rather large 
incompressive $\text{Pi-D}^s$ contributions. 
Alternatively, it is possible that highly compressive flows with very small solenoidal velocity might exhibit significant $\text{Pi-D}$ that is of almost fully compressive origin, being dominated by $\text{Pi-D}^c$. This could be for example a flow consisting of compressional waves and shocklets.
Allowing for such possibilities, 
one should, in general,
interpret $\text{Pi-D}$ as ``incompressive'' with great caution. We understand of course that having a full Helmholtz decomposition of the flow would alleviate both of these ambiguities, but this is not available with MMS data in spite of its prodigious capabilities.

\begin{acknowledgments}
This research is partially supported by the MMS Theory, Modeling and Data Analysis team under NASA grant 80NSSC19K0565, 
by the NASA LWS program under grants 80NSSC20K0198 and 
80NSSC22K1020, 
and a subcontract from the New Mexico consortium 655-001, 
a NASA Heliophysics MMS-GI grant through a Princeton subcontract SUB0000517, and by the National Science Foundation
Solar Terrestrial Program grant 
AGS-2108834. 
Y.Y. is supported by 2024 Ralph E. Powe Junior Faculty Enhancement Award and the University of Delaware General University Research Program grant.
This research was also supported by the International Space Science Institute (ISSI) in Bern, through ISSI International Team projects \#556 (Cross-scale energy transfer in space plasmas) and \#23-588 (Unveiling energy conversion and dissipation in nonequilibrium space plasmas).
We would like to acknowledge high-performance computing support from Cheyenne (doi:10.5065/D6RX99HX) and Derecho (https://doi.org/10.5065/qx9a-pg09) provided by NCAR's Computational and Information Systems Laboratory, sponsored by the National Science Foundation.

\end{acknowledgments}

\bibliography{main}

\end{document}